\documentclass[english,american,9pt,conference,epsfig]{IEEEtran}
\usepackage[T1]{fontenc}
\usepackage[latin9]{inputenc}
\usepackage{float}
\usepackage{amsmath}
\usepackage{graphicx}

\makeatletter
\@ifundefined{date}{}{\date{}}

\usepackage{pstricks, pst-node, pst-plot, pst-circ}
\usepackage{moredefs}

\usepackage{psfrag}

\usepackage{amsmath,graphicx}

\usepackage{multirow}
\usepackage{booktabs}

\usepackage{times,amsmath,epsfig}
\usepackage{epstopdf}
\usepackage{rotating}

\newcommand{\HP}{\text{H\hspace{-0.25mm}P}}
\newcommand{\LP}{\text{L\hspace{-0.25mm}P}}
\newcommand{\sigf}{f}
\newcommand{\LL}{\text{L\hspace{-0.25mm}L}}
\newcommand{\LH}{\text{L\hspace{-0.25mm}H}}
\newcommand{\HL}{\text{H\hspace{-0.25mm}L}}
\newcommand{\HH}{\text{H\hspace{-0.25mm}H}}
\newcommand{\pred}{p}

\newcommand{\fig}{Fig.}
\newcommand{\tab}{Table}

\newcommand{\Fig}[1]{\fig{}~#1}
\newcommand{\Tab}[1]{\tab{}~#1}

\newcommand{\Legall}{LeGall~5/3 wavelet}

\newcommand{\jptwok}{JPEG~2000}

\newcommand{\eind}{\mbox{1-D}}
\newcommand{\zweid}{\mbox{2-D}}
\newcommand{\dreid}{\mbox{3-D}}
\newcommand{\dreidt}{\mbox{3-D+t}}

\newcommand{\blockbased}{block-based}

\newcommand{\spaceBelowFig}{\vspace{-4mm}}

\newcommand{\spaceBeforeCapt}{\vspace{-1mm}}

\author{
{Wolfgang Schnurrer, Tobias Tröger, Thomas Richter, Jürgen Seiler, and André Kaup}
\vspace{1.6mm}\\ 
\fontsize{10}{10}\selectfont\itshape 
Multimedia Communications and Signal Processing\\
Friedrich-Alexander Universität Erlangen-Nürnberg (FAU), Cauerstr. 7, 91058 Erlangen, Germany
\fontsize{9}{9}\selectfont\ttfamily\upshape

\vspace{1.2mm}\\ 
\fontsize{10}{10}\selectfont\rmfamily\itshape 

\fontsize{9}{9}\selectfont\ttfamily\upshape 
Email: \{ schnurrer, troeger, richter, seiler, kaup \} @lnt.de
}

\makeatother

\usepackage{babel}
\begin{document}

\title{Efficient Lossless Coding of Highpass Bands from Block-based Motion
Compensated Wavelet Lifting Using \jptwok{}}

\maketitle

\selectlanguage{english}%

\selectlanguage{american}%

\maketitle

\begin{figure}[b] \parbox{\hsize}{
IEEE VCIP'14, Dec. 7 - Dec. 10, 2014, Valletta, Malta.

978-1-4799-6139-9/14/\$31.00 \ \copyright 2014 IEEE. 
}\end{figure}
\begin{abstract}
Lossless image coding is a crucial task especially in the medical
area, e.g., for volumes from  Computed Tomography or Magnetic Resonance
Tomography. Besides lossless coding, compensated wavelet lifting offers
a scalable representation of such huge volumes. While compensation
methods increase the details in the lowpass band, they also vary
the characteristics of the wavelet coefficients, so an adaption of
the coefficient coder should be considered. We propose a simple invertible
extension for \jptwok{} that can reduce the filesize for lossless
coding of the highpass band by 0.8\% on average with peak rate saving
of 1.1\%.
\end{abstract}

\section{\label{sec:Introduction}Introduction}

Multi-dimensional data volumes, like \dreid{} or \dreidt{} image
data from Computed Tomography (CT) or Magnetic Resonance Tomography,
can become unhandy large very fast. Storing, transmitting, processing
and even displaying such huge volumes becomes a challenging task.
A scalable representation is desired, where a coarse representation
can be used for previewing or fast browsing, while interesting  areas
can be reconstructed lossless \cite{calderbank1998wavelet}. The latter
is very important, e.g., for diagnosis in telemedical applications. 

\begin{figure}
\psfragscanon

\psfrag{time}{$t$}
\psfrag{comp}{compensated}
\psfrag{W}{Wavelet}
\psfrag{L}{Lifting}
\psfrag{hp}{$\HP_t$}
\psfrag{lp}{$\LP_t$}
\psfrag{dwt2}{\zweid{} WT in}
\psfrag{xy}{$xy$-direction}
\psfrag{x}{{\tiny $x$}}
\psfrag{y}{{\tiny $y$}}

\includegraphics[width=0.99\columnwidth]{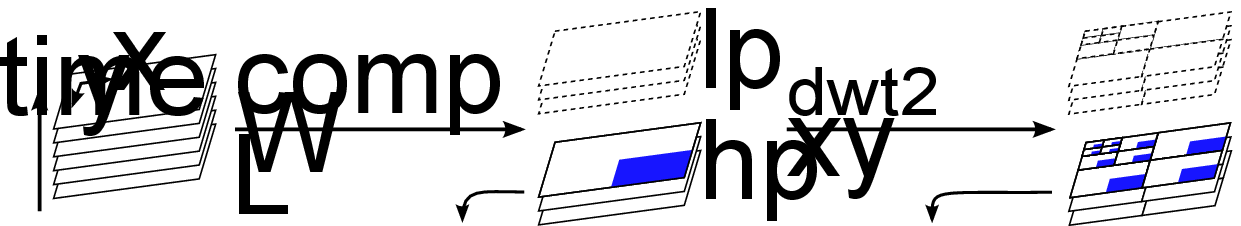}

\psfragscanoff

\includegraphics[width=0.49\columnwidth]{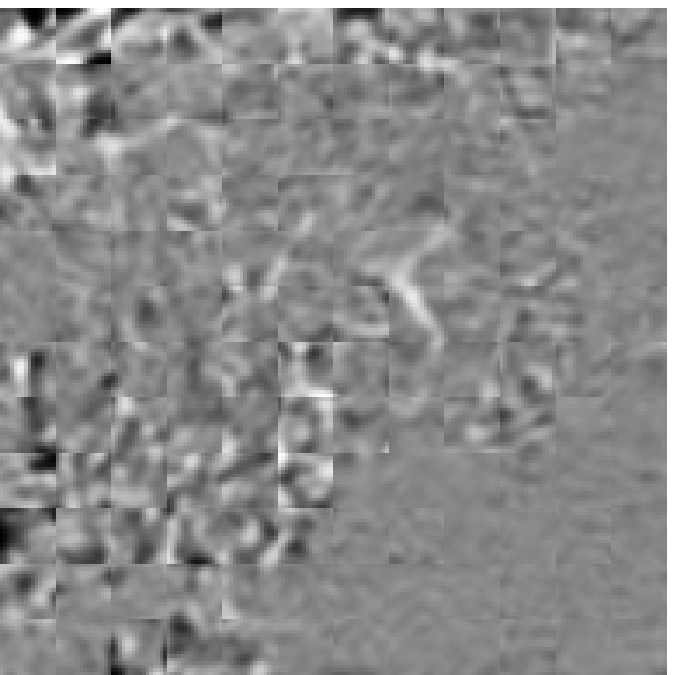}\hfill{}\includegraphics[width=0.49\columnwidth]{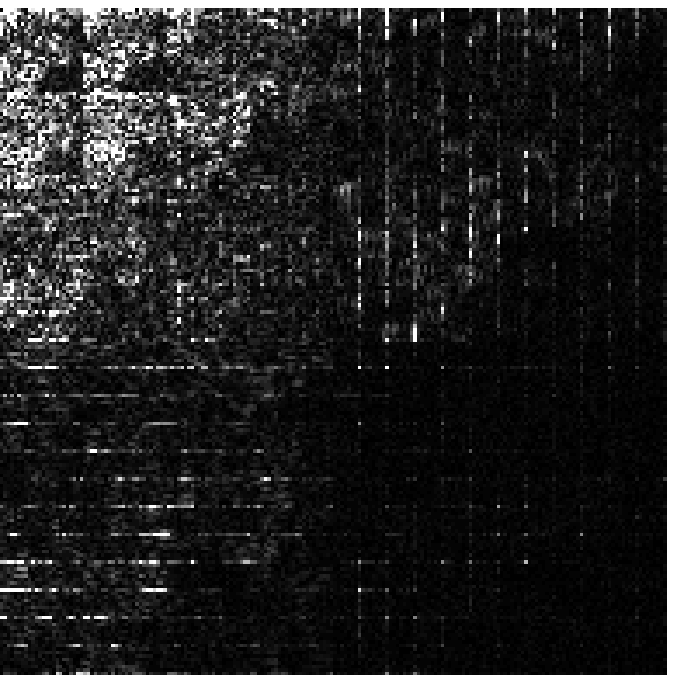}

\spaceBeforeCapt{}

\caption{\label{fig:Wavelet-coefficients}Visualization of the occurring structures
in the highpass. The marked details from the block-diagram are shown
below. Left: detail of the highpass band from wavelet lifting with
block-based compensation (gray=0), Right: corresponding further decomposition
in $xy$-direction (absolute values)}
\spaceBelowFig{}
\end{figure}

For a multi-dimensional wavelet transform (WT), the \eind{} WT is
applied successively along the different dimensions, shown in \Fig{\ref{fig:Wavelet-coefficients}}.
The lowpass band of a WT can be considered as downscaled version of
the original signal. In contrast to a subsampling when every other
frame is taken, the lowpass band contains information from the complete
original signal. To obtain a more detailed lowpass band, the WT in
temporal or $z$-direction can be extended by compensation methods
\cite{garbasTCSVT,schnurrer2012vcip,secker2003}. \Fig{\ref{fig:Wavelet-coefficients}}
shows occurring structures in the highpass band caused by block-based
compensation. Deformable motion models \cite{secker2003,sullivan1991,Weinlich2012}
can avoid these structures. Since these models usually use a complex
iterative estimation process, this paper focuses on \blockbased{}
compensation. However, the characteristics of the WT coefficients
are varied significantly by the \blockbased{} compensation, as also
shown in \cite{Abhayaratne2000}. Several methods exist for coding
wavelet coefficients. They all exploit characteristics of the coefficients
of a traditional WT, i.e., without a compensation method incorporated.
Extensions like a variable blocksize \cite{hevc_overview} can be
used but the shown structures can still occur. We observed that the
lossless coding efficiency can be improved when the coding method
is adapted to the variation of the coefficients in compensated lifting. 

In \cite{Abhayaratne2000}, this problem is addressed by adapting
the wavelet basis to the characteristics of the signal. The highpass
band of compensated wavelet lifting can be considered as prediction
residual as well. Instead of modifying the wavelet basis, we present
a different approach that just adapts the order of the coefficients
prior to the entropy coder. Our goal is to keep the coding method
unchanged, so specialized hardware coders can still be used.

We propose a method to improve the efficiency for lossless coding
of highpass coefficients of a WT with block-based compensation using
\jptwok{} \cite{christopoulos2002,taubman2000,Taubman200249}. \jptwok{}
is a wavelet-based image coding method that is also part of the DICOM
standard \cite{dicom2012}. We present a re-sorting of the compensated
highpass coefficients that can also be implemented as a preprocessing
step of a standard \jptwok{} coder. This paper focuses on the computation
and the processing of the highpass band. An efficient processing of
the lowpass band has already been proposed in \cite{schnurrer2013}.
\begin{figure*}[!t]
\begin{center}
\psfragscanon
\psfrag{dwt}{\zweid{} WT in}
\psfrag{xy}{$xy$-direction}
\psfrag{cblock}{coding block}
\psfrag{cd}{coding}
\psfrag{bl}{block}
\psfrag{T1}{Tier 1}
\psfrag{T2}{Tier 2}
\psfrag{ebcot}{EBCOT}
\psfrag{bs}{bitstream}
\psfrag{inp}{input image}
\psfrag{wc}{wavelet coefficients}
\psfrag{wa}{wavelet}
\psfrag{co}{coefficients}
\psfrag{t2a}{optimize order}
\psfrag{t2b}{of embedded}
\psfrag{t2c}{bitstreams}

\psfrag{x}{$x$}
\psfrag{y}{$y$}

\includegraphics[height=3cm]{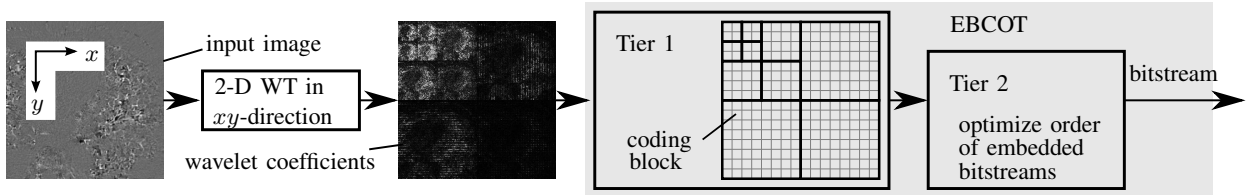}

\psfragscanoff
\end{center}

\spaceBeforeCapt{}\spaceBeforeCapt{}\caption{\label{fig:jptwok-chain}Simplified processing chain of \jptwok{},
according to \cite{christopoulos2002,taubman2000,Taubman200249}}
\spaceBelowFig{}
\end{figure*}

In Section~\ref{sec:Compensated-Wavelet-Lifting}, we briefly review
compensated wavelet lifting and sketch the coding chain of \jptwok{}.
In Section~\ref{sec:Coefficient-Re-sorting} we introduce our coefficient
re-sorting approach into the coding framework together with an optimum
as well as a low complexity decision approach. Simulation results
and discussion follow in Section~\ref{sec:Simulation}. Section~\ref{sec:Conclusion}
concludes this paper.

\section{Compensated Wavelet Lifting}

\label{sec:Compensated-Wavelet-Lifting}Wavelet lifting is an efficient
implementation of a wavelet transform (WT) \cite{daubechies1998}.
A WT can be applied in temporal direction to obtain temporal scalability.
To reduce the motion artifacts and ghosting artifacts in the lowpass
band for a better quality, motion compensation methods can be implemented
directly into the transform \cite{garbasTCSVT}. Therefore, the compensated
frames $\pred_{2t-1}$ and \foreignlanguage{english}{$\pred_{2t+1}$}
are subtracted from the current frame $\sigf_{2t}$ to compute the
highpass frame $\HP_{t}$ with index $t$ as shown in \eqref{eq:H-L53}
for the \Legall{}.

\selectlanguage{english}%
\begin{equation}
\HP_{t}=\sigf_{2t}-\left\lfloor \frac{1}{2}\left(\pred_{2t-1}+\pred_{2t+1}\right)\right\rfloor \label{eq:H-L53}
\end{equation}
\foreignlanguage{american}{This further leads to a reduction of the
energy in the highpass band and thus a better decorrelation of the
signal and higher transform gain \cite{schnurrer2012vcip}. }

\selectlanguage{american}%
The resulting subbands from the compensated transform are coded frame
by frame with \jptwok{}. \jptwok{} is a wavelet-based image coder
and fits seamless into a wavelet-based framework. \Fig{\ref{fig:jptwok-chain}}
shows a simplified processing chain of \jptwok{}. An input image
is decomposed using a \zweid{} WT. The coefficients are then coded
using Embedded Block Coding with Optimized Truncation (EBCOT) \cite{Taubman200249}.
Therefore, the subbands are subdivided into coding blocks. EBCOT consists
of two tiers. In Tier~1, the coefficients of each coding block are
traversed in a specific scan order and arithmetically coded into an
embedded bitstream. Tier~2 operates on the results of Tier~1 and
determines the optimum order of the embedded bitstreams, i.e., the
coding blocks, in the resulting final bitstream for optimum scalability.
For a more detailed description, please refer to \cite{christopoulos2002,taubman2000,Taubman200249}.

To summarize, all subbands are processed independently by \jptwok{}.
After Tier~1, the rate needed for each subband can be computed by
summing up the lengths of all embedded bitstreams. The next section
describes our proposed method making use of these coder properties
for adapting the characteristics of compensated highpass frames to
increase the coding efficiency of \jptwok{}.

\section{Proposed Coefficient Re-sorting}

\label{sec:Coefficient-Re-sorting}Block-based compensation methods
can lead to a predictor containing block structures, especially when
the translatory motion model does not exactly fit the occurring motion.
The highpass band can be regarded as prediction error signal when
a compensated transform is considered. The block structures in the
highpass band also have to be coded. This can increase the amount
of bits needed for coding \cite{Abhayaratne2000}.

Neighboring pixels in the highpass band are still correlated, so a
further decomposition in $xy$-direction is reasonable. We observed
that the decomposition of a highpass frame with block structures leads
to characteristic structures that are dependent on the block-size
of the block-based compensation. These structures are shown in \Fig{\ref{fig:Wavelet-coefficients}}
on the right.

\begin{figure}[H]
\begin{center}
\psfragscanon
\psfrag{LL}{{\small $\LL_4$ }}
\psfrag{LH4}{{\small $\LH_4$ }}
\psfrag{HL4}{{\small $\HL_4$ }}
\psfrag{HH4}{{\small $\HH_4$ }}

\psfrag{LH3}{{\small $\LH_3$ }}
\psfrag{HL3}{{\small $\HL_3$ }}
\psfrag{HH3}{{\small $\HH_3$ }}

\psfrag{LH2}{$\LH_2$}
\psfrag{HL2}{$\HL_2$}
\psfrag{HH2}{$\HH_2$}

\psfrag{LH1}{$\LH_1$}
\psfrag{HL1}{$\HL_1$}
\psfrag{HH1}{$\HH_1$}

\includegraphics[height=3.3cm]{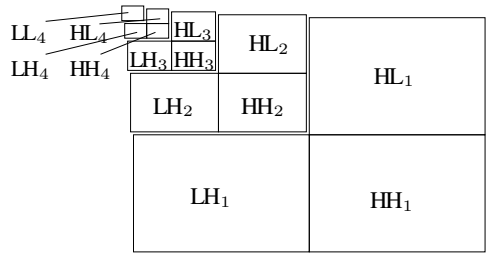}

\psfragscanoff
\end{center}

\spaceBeforeCapt{}\spaceBeforeCapt{}

\caption{\label{fig:wavelet-decomposition} Notation of the subbands of a dyadic
2-D wavelet decomposition with four decompositions}
\end{figure}

The first wavelet decomposition yields four subbands, namely $\LL_{1}$,
$\HL_{1}$, $\LH_{1}$, and $\HH_{1}$. \Fig{\ref{fig:wavelet-decomposition}}
shows a dyadic decomposition with four steps, where a further decomposition
of the lowpass band $\LL_{i}$ leads to the subbands $\LL_{i+1}$,
$\HL_{i+1}$, $\LH_{i+1}$ and $\HH_{i+1}$. For lossless coding,
the fully reversible integer \Legall{} \cite{calderbank1998wavelet}
is used for the decomposition in the $xy$-direction \cite{christopoulos2002}.

The coefficients in the $\LH$ bands correspond to horizontal edges,
i.e., high frequencies in vertical direction and coefficients in the
$\HL$ bands correspond to vertical edges, i.e., high frequencies
in horizontal direction. The horizontal respectively vertical edges
from block boundaries change the characteristic of the coefficients
significantly. The entropy coder of \jptwok{} is not able to exploit
the occurring structures because only a small local neighborhood of
coefficients is used for prediction \cite{Taubman200249}.

\begin{figure*}
\psfragscanon

\psfrag{Hband}{frame of the compensated \HP{}-band}

\psfrag{dwt2d}{\zweid{} WT in}
\psfrag{xy}{$xy$-direction}
\psfrag{025bs}{$\frac{1}{4}bs$}
\psfrag{05bs}{$\frac{1}{2}bs$}
\psfrag{bs}{$bs$}

\psfrag{rsrt}{\begin{rotate}{90}re-sort all subbands\end{rotate}}
\psfrag{asb}{\begin{rotate}{90}all subbands\end{rotate}}

\psfrag{col}{column of}
\psfrag{row}{row of}
\psfrag{rows}{rows of}
\psfrag{neigh}{neighboring}
\psfrag{coeff}{coefficients}

\psfrag{rps}{rates per subband}
\psfrag{jp2kt1}{\jptwok{} Tier 1}
\psfrag{jp2kt2}{\jptwok{} Tier 2}
\psfrag{com1}{compare rates and}
\psfrag{com2}{decide for smaler}

\psfrag{comth1}{compare to thresholds}
\psfrag{comth2}{re-sort subband if smaller}

\psfrag{lcdec}{\begin{rotate}{90}low complexity (LC) decision\end{rotate}}
\psfrag{optdec}{\begin{rotate}{90}optimum (OPT) decision\end{rotate}}

\psfrag{lcres}{result from low complexity (LC) decision}
\psfrag{optres}{result from optimum (OPT) decision}
\psfrag{trdres}{trad. \jptwok{}}

\psfrag{thhh1}{0.3}
\psfrag{thhh2}{0.3}
\psfrag{thhh3}{0.6}

\psfrag{thlh1}{0.5}
\psfrag{thlh2}{0.6}
\psfrag{thlh3}{0.6}

\psfrag{thhl1}{0.5}
\psfrag{thhl2}{0.6}
\psfrag{thhl3}{0.6}

\includegraphics[width=1\textwidth]{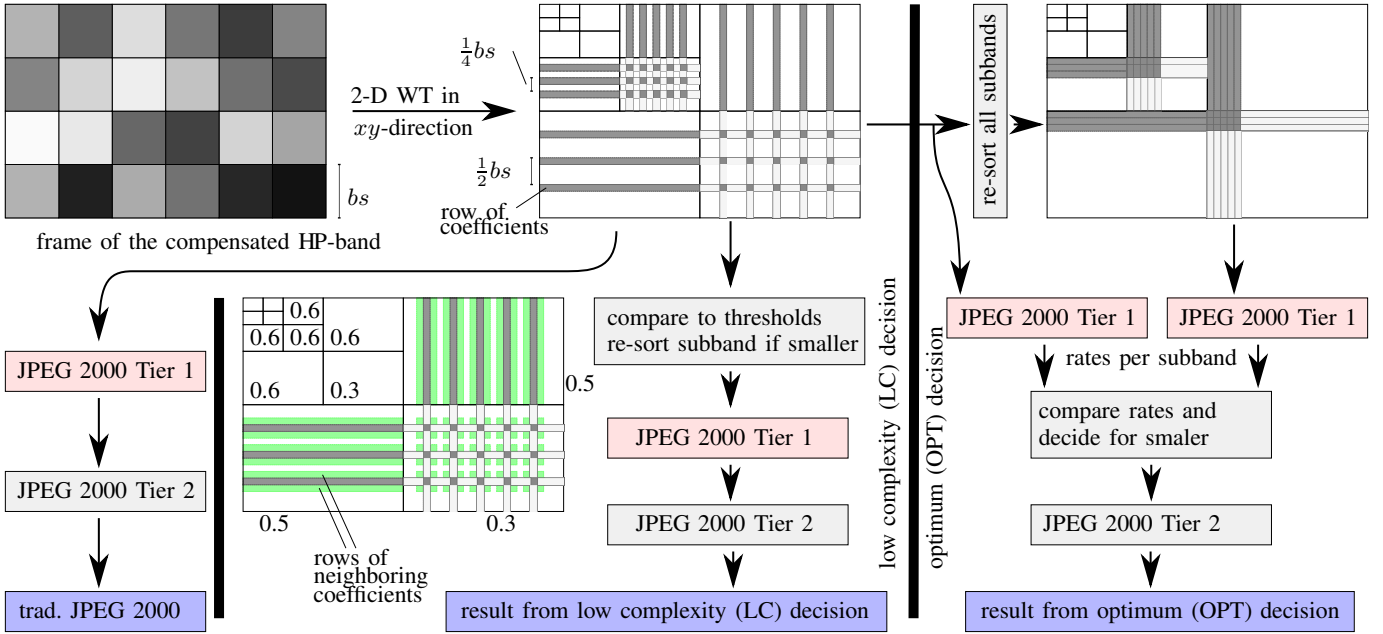}

\psfragscanoff
\spaceBeforeCapt{}

\caption{\label{fig:Block-diagram}Block diagram showing our proposed coefficient
re-sorting algorithm with optimum (OPT) decision approach on the right
and low complexity (LC) decision approach in the center. For comparison,
traditional \jptwok{} is shown on the left.}
\spaceBelowFig{}
\end{figure*}
\Fig{\ref{fig:Block-diagram}} shows our proposed re-sorting algorithm
and our two decision approaches. The occurring structures are represented
by gray color in the top center resulting from the further decomposition
of the compensated highpass band, shown on the left. To exploit these
structures, we propose a re-sorting of the coefficients. In the subbands
of the first decomposition in $xy$-direction, namely $\HL_{1}$,
$\LH_{1}$, and $\HH_{1}$, the distance between the structures is
one half of the blocksize $bs$ of the compensation method. For the
second decomposition the distance is $\frac{1}{4}bs$, as shown in
\Fig{\ref{fig:Block-diagram}}. For a blocksize $bs$ of $16\times16$
pixels, the distance is 8 in $\HL_{1}$, $\LH_{1}$, and $\HH_{1}$,
4 in $\HL_{2}$, $\LH_{2}$, and $\HH_{2}$ and 2 in $\HL_{3}$, $\LH_{3}$,
and $\HH_{3}$. In the fourth decomposition, the structures are next
to each other and thus within the reach of the internal predictor
of EBCOT \cite{Taubman200249}. The maximum number of decompositions
$d_{m}$ for re-sorting to be evaluated computes to 
\begin{equation}
d_{m}=\log_{2}\left(bs\right)-1.\label{eq:max-number-decomp}
\end{equation}

The re-sorting works as follows: for the $\LH$ bands, all rows of
coefficients corresponding to block boundaries are moved to the top,
as illustrated on the top right in \Fig{\ref{fig:Block-diagram}}.
For the $\HL$ bands, all respective columns are moved to the left.
For the $\HH$ bands, both operations are applied. The result is shown
in \Fig{\ref{fig:Block-diagram}} on the top right. Please note that
the coefficients are re-sorted and the order of the code-blocks is
not modified, i.e., the two tiers remain unchanged.

On the right side of \Fig{\ref{fig:Block-diagram}}, the algorithm
for obtaining the optimum decision is shown. The coefficients per
subband can be modified and it can be checked whether the rate decreases.
For each subband, the rate needed for traditional coding, i.e., standard
\jptwok{} without re-sorting, as well as the rate needed for coding
the re-sorted coefficients is determined by executing the Tier~1
coding pass. Next, for each subband, the smaller rate corresponds
to the optimum decision.

Due to arithmetic coding, Tier~1 is a quite complex part of \jptwok{}.
Executing Tier~1 twice increases the computational complexity a lot.
To avoid this, a simple decision method was developed, shown in the
bottom center of \Fig{\ref{fig:Block-diagram}}. After the wavelet
decomposition, a quotient is computed for every subband. Therefore,
for every subband, the sum of the absolute values of the coefficients
corresponding to block boundaries (gray color) is computed in a first
step. Next, the sum of the absolute values of the coefficients corresponding
to the neighboring coefficients (green color) is computed. Then, the
quotient of the previously computed two values is compared to a threshold.
When the quotient is small enough, i.e., the difference between the
block boundary coefficients and their neighbors is big enough, the
coefficients of the subband are re-sorted. For the $\HL$ bands, the
neighbors (green) left and right of each gray column are summed up.
So the values of the gray columns are multiplied by 2 to compensate
for the twice as many neighbors. This is done analogue for the respective
rows of the $\LH$ bands. For the $\HH$ bands, the absolute values
of the four diagonal neighbors of each dark gray coefficient are summed
up and the absolute sum of the dark gray coefficients is multiplied
by 4 respectively. As shown in \Fig{\ref{fig:Block-diagram}}, the
decision is made before Tier~1 for this low complexity approach,
so Tier~1 is executed only once.

The traditional \jptwok{} processing chain is again shown on the
left side for comparison as well as to show all cases of our simulation
setup in \Fig{\ref{fig:Block-diagram}}.

For signaling the decision to the decoder, one additional bit for
each subband is needed. One more bit per frame indicates whether re-sorting
is used at all. If the re-sorting is not used, the overall filesize
will increase only by one bit per frame. The operations are all reversible
so the property of lossless coding is not harmed.

The re-sorting can be implemented as a preprocessing step before \jptwok{}-encoding
and a postprocessing step after \jptwok{}-decoding, so a standard
\jptwok{} coder can be used. For the post-processing, the wavelet
decomposition has to be computed after \jptwok{} decoding, followed
by the inverse coefficient re-sorting and an inverse WT.

\section{Simulation Results}

\label{sec:Simulation}For evaluating our method, we used different
CT data sets. One \dreidt{} CT \textit{heart} data set\foreignlanguage{english}{}%
\footnote{The CT volume data set was kindly provided by Siemens Healthcare.%
} was used where the transform is applied in slice-direction (\textit{heart
spat}) and in time-direction (\textit{heart time}). Further, we tested
four \dreid{} CT \textit{head} data sets and four \dreid{} CT \textit{thorax}
data sets\foreignlanguage{english}{}%
\footnote{The CT volume data sets were kindly provided by Prof. Dr. med. Dr.
rer. nat. Reinhard Loose from the Klinikum Nürnberg Nord.%
}. 

We applied a compensated \Legall{} in temporal, respectively slice-direction
and evaluated the lossless coding of the highpass coefficients. The
block-based compensation was used with a blocksize of $16\times16$
with a full-search within a search range of 15. The resulting highpass
bands were coded frame by frame using the \jptwok{} implementation
\cite{openjpeg} with 7 wavelet decompositions. The re-sorting was
evaluated for the subbands from the first 3 decompositions in $xy$-direction,
as computed by \eqref{eq:max-number-decomp}.

\begin{table*}[!t]
\begin{center}
\begin{tabular}{|c|c|c|c|c|c|c|}

\toprule

\multirow{2}{*}{sequence} & \multicolumn{3}{c|}{filesize in bytes} & \multicolumn{3}{c|}{savings } \tabularnewline
 & trad. \jptwok{} & re-sort OPT & re-sort LC & abs OPT & abs LC & rel LC/OPT in \% \tabularnewline

\cmidrule{1-7}


 \it{heart spat} & 98880861 & \textbf{97776086} & \textbf{97776578} &  1104775 &  1104283 &  -0.045  \tabularnewline 
 \it{heart time} & 92972210 & \textbf{92122327} & \textbf{92122485} &  849883 &  849725 &  -0.019  \tabularnewline 
 \it{head} & \textbf{3751832} & 3751834 & 3751834 &  -2 &  -2 &  0  \tabularnewline 
 \it{head 2} & \textbf{7318102} & 7318107 & 7318107 &  -5 &  -5 &  0  \tabularnewline 
 \it{head 3} & 1502434 & \textbf{1492997} & \textbf{1493900} &  9437 &  8534 &  -9.569  \tabularnewline 
 \it{head 4} & 1814188 & \textbf{1809398} & \textbf{1811464} &  4790 &  2724 &  -43.132  \tabularnewline 
 \it{thorax 1} & 7661846 & \textbf{7651164} & \textbf{7651651} &  10682 &  10195 &  -4.559  \tabularnewline 
 \it{thorax 2} & \textbf{5979688} & 5979693 & 5979693 &  -5 &  -5 &  0  \tabularnewline 
 \it{thorax 3} & \textbf{7850487} & 7850492 & 7850492 &  -5 &  -5 &  0  \tabularnewline 
 \it{thorax 4} & 9164221 & \textbf{9163155} & \textbf{9163335} &  1066 &  886 &  -16.886 
 \\

\cmidrule{1-7}
average & & $-0.836 \%$ & $-0.834 \%$ & & & \tabularnewline
\cmidrule{1-7}
\end{tabular}
\end{center}

\caption{\label{tab:LOW-COMPLEXITY-results}Coding results for traditional
\jptwok{} and our proposed re-sorting algorithm with optimum (OPT)
decision approach and low complexity (LC) decision approach}
\end{table*}
\Tab{\ref{tab:LOW-COMPLEXITY-results}} shows the lossless coding
results for the compensated highpass coefficients using the three
cases shown in \Fig{\ref{fig:Block-diagram}}, namely traditional,
i.e., standard \jptwok{}, and the proposed re-sorting method with
optimum (OPT) and low complexity (LC) decision. The thresholds for
the LC approach are given in \Fig{\ref{fig:Block-diagram}}. The overhead
information for signaling the re-sorting is included.

The absolute savings in bytes are given for the two re-sorting approaches
against traditional \jptwok{}. Negative values indicate that more
data has to be stored using our proposed method due to signaling overhead,
e.g., for \textit{head2}, 36 \HP{} frames result in an overhead of
$\left\lceil \log_{2}36\right\rceil =5$~bytes. For medical CT volumes,
the proposed re-sorting can reduce the number of bits for lossless
coding by 0.8\% on average with peak rate saving of 1.1\%. This is
notable since in general, even small gains are hard to achieve in
lossless coding. Compared to the achievable gains, the loss due to
the signaling information is negligible. The column on the right compares
the results from the two decision approaches showing that the LC decision
performs mostly close to the OPT decision. Although the gains are
a little smaller, the LC approach achieves gains where OPT performs
better then traditional \jptwok{}.

The achievable gain strongly depends on the content of the sequence.
The re-sorting can be applied to video sequences as well resulting
in smaller gains. The medical sequences show less high frequency content
compared to the video sequences. We observed, that if the absolute
values of the coefficients corresponding to the block boundaries are
significantly larger than the surrounding neighboring coefficients
it is advantageous to re-sort the coefficients to achieve a higher
compression.

As shown in \Fig{\ref{fig:Block-diagram}}, the optimum decision needs
to run the Tier~1 part two times, which then leads to the optimum
results. Our low complexity decision approach shows that this increase
of the encoder complexity can be avoided by a decision method, that
determines more efficiently whether it is advantageous to apply the
re-sorting for a subband. Furthermore, the decoder complexity is only
changed marginally as only a simple re-ordering of the coefficients
is necessary.

\section{Conclusion}

\label{sec:Conclusion}In this paper we propose an efficient method
that can improve lossless compression of highpass bands from block-based
compensated wavelet lifting of medical CT data sets using \jptwok{}.
We showed that an adaption of the compensated coefficients to the
coder can improve the coding efficiency. The proposed reversible method
can be implemented as preprocessing before encoding and postprocessing
after decoding, so a standard \jptwok{} encoder and decoder can be
used. Within our simulation data set, the filesize of the lossless
coded highpass band was reduced by 0.8\% on average with peak rate
saving of 1.1\%. The optimum decision performs best but has a high
computational complexity. Our proposed low complexity decision approach
comparing sums of coefficients performs close to the optimum decision.

The proposed re-sorting method is not limited to highpass bands from
compensated wavelet lifting but can be applied to wavelet-based coding
of residuals from block-based motion compensation as well. Further
work aims at an evaluation of the lossy-to-lossless scalability as
well as a detailed complexity analysis.

\section*{Acknowledgment}

We gratefully acknowledge that this work has been supported by the
Deutsche Forschungsgemeinschaft (DFG) under contract number KA~926/4-2.

\bibliographystyle{IEEEbib}
\bibliography{bib/bibliography}

\end{document}